\documentclass[%
 aip,
 amsmath,amssymb,
 reprint,%
]{revtex4-1}

\usepackage{graphicx}
\usepackage{dcolumn}
\usepackage{bm}

\usepackage[utf8]{inputenc}
\usepackage[T1]{fontenc}
\usepackage{mathptmx}
\usepackage{etoolbox}
\usepackage{siunitx}
\usepackage{amsmath,amssymb,amsfonts}%
\usepackage{amsthm}%
\usepackage{mathrsfs}%
\usepackage{hyperref}

\makeatletter
\def\@email#1#2{%
 \endgroup
 \patchcmd{\titleblock@produce}
  {\frontmatter@RRAPformat}
  {\frontmatter@RRAPformat{\produce@RRAP{*#1\href{mailto:#2}{#2}}}\frontmatter@RRAPformat}
  {}{}
}%
\makeatother
\begin{document}

\preprint{AIP/123-QED}

\title[Inverse Design of Polaritonic Devices]{Inverse Design of Polaritonic Devices}
\author{Oliver Kuster}
 \affiliation{Institute of Theoretical Solid State Physics, Karlsruhe Institute of Technology (KIT), Kaiserstrasse 12, 76131 Karlsruhe, Germany.}
 \email{oliver.kuster@kit.edu}

\author{Yannick Augenstein}%
 \affiliation{Institute of Theoretical Solid State Physics, Karlsruhe Institute of Technology (KIT), Kaiserstrasse 12, 76131 Karlsruhe, Germany.}

\author{Carsten Rockstuhl}
 \affiliation{Institute of Theoretical Solid State Physics, Karlsruhe Institute of Technology (KIT), Kaiserstrasse 12, 76131 Karlsruhe, Germany.}
 \affiliation{Institute of Nanotechnology, Karlsruhe Institute of Technology (KIT), Kaiserstrasse 12, 76131 Karlsruhe, Germany.}

\author{Thomas Jebb Sturges}%
 \affiliation{Institute of Theoretical Solid State Physics, Karlsruhe Institute of Technology (KIT), Kaiserstrasse 12, 76131 Karlsruhe, Germany.}

\date{\today}

\begin{abstract}
Polaritons, arising from the strong coupling between excitons and photons within microcavities, hold promise for optoelectronic and all-optical devices. They have found applications in various domains, including low-threshold lasers and quantum information processing. To realize complex functionalities, non-intuitive designs for polaritonic devices are required. 
In this contribution, we use finite-difference time-domain simulations of the dissipative Gross-Pitaevskii equation, written in a differentiable manner, and combine it with an adjoint formulation. Such a method allows us to use topology optimization to engineer the potential landscape experienced by polariton condensates to tailor its characteristics on demand. The potential directly translates to a blueprint for a functional device, and various fabrication and optical control techniques can experimentally realize it. We inverse-design a selection of polaritonic devices, i.e., a structure that spatially shapes the polaritons into a flat-top distribution, a metalens that focuses a polariton, and a nonlinearly activated isolator. The functionalities are preserved when employing realistic fabrication constraints such as minimum feature size and discretization of the potential. Our results demonstrate the utility of inverse design techniques for polaritonic devices, providing a stepping stone toward future research in optimizing systems with complex light-matter interactions.
\end{abstract}

\maketitle
Topology optimization \cite{bendsoe1988generating, Bendsoe2003Topology}, a powerful tool for inverse design, has revolutionized the creation of structures and devices across various domains of physics and engineering. By leveraging gradient-based algorithms, topology optimization iteratively optimizes a predefined design region towards a structure with tailored functionality. Remarkably, with this method, one can calculate all gradients exactly, with just two solves of the system equations, namely a forward simulation and a ``backward'' adjoint simulation, regardless of the number of design parameters \cite{Jensen2011Topology}. That efficiency enables practically unlimited free-form optimization of a design region. Topology optimization has been used in mechanical engineering for decades and is integrated into commercial design tools \cite{Christiansen2021Inverse}. Since then, topology optimization found its way into many branches of engineering and science. In the last decade, the photonics community has begun to widely use such inverse design techniques \cite{Christiansen2021Inverse, Molesky2018Inverse, Jensen2011Topology}. This approach, facilitated by fully differentiable photonic solvers featuring built-in adjoint solvers \cite{Hammond2022High, Oskooi2010MEEP}, has led to the design of diverse devices such as multiplexers \cite{Piggott2015inverse}, metalenses \cite{Christiansen2020Fullwave}, mode converters \cite{Augenstein2020Inverse} and many others \cite{Augenstein2018Inverse, Augenstein2022Inverse, Gladyshev2023Inverse}. Importantly, topology optimization readily incorporates fabrication constraints through direct inclusion or soft constraints in the optimization process. This work extends the applicability range and explores topology optimization to design polariton devices.

Polaritons are quasiparticles that arise from the strong light-matter coupling between excitons in quantum wells or wires and confined photonic modes in optical microcavities \cite{Hopfield1958Theory, kavokin2017microcavities}. Their bosonic nature allows a coherent state in which a single mode is macroscopically occupied \cite{Byrnes2014Exciton}. Polaritons exhibit beneficial properties for diverse applications such as low-threshold lasers and light emitters \cite{Das2011Room}, polariton optoelectronic circuitry and logic gates \cite{Liew2008Optical}, and quantum information processing \cite{Angelakis2017Quantum, Kavokin2022Polariton}. Examples of polariton devices include waveguide couplers \cite{Beierlein2021Propagative}, transistors \cite{Zasedatelev2019Room}, and directional antennae \cite{Aristov2023Directional}. Central to the design of polariton devices is the engineering of the effective potential landscape experienced by polaritons \cite{Schneider2017Exciton}. Various approaches, such as etched mesa structures \cite{Kaitouni2006Engineering}, local variations of the cavity height \cite{Winkler2015Polariton, Adiyatullin2017Periodic}, electrostatic straps \cite{Gartner2007Micropatterned}, and excitonic reservoir confinement mechanisms \cite{Schmutzler2015All, Wertz2010Spontaneous}, enable control over the polariton potential landscape. Often, a description based on a generalized open-dissipative Gross-Pitaevskii equation is enough to model the dynamics of the system. This is a widely used approach in both inorganic \cite{kavokin2017microcavities} and organic \cite{Keeling2020Boseeinstein} microcavities. In this model, the potential landscape felt by the polaritons is included in the equations as a spatially dependent scalar field that multiplies the polariton condensate wavefunction. One can consider the effective potential as a quantity designed to implement specific functionalities within the constraints imposed by the particular physical system.

In this article, we use topology optimization to engineer the potential landscape of polaritons, optimizing for specific polariton devices. In this case, the discretized values of the scalar field are the design variables. As discussed, the adjoint method allows us to optimize each point of the potential landscape independently in a free-form manner with just two solves of the system equations. Due to the nonlinearity in the equations, we use here finite-difference time-domain simulations to solve for the dynamics (or the steady state) of the polaritons. We also consider a specific set of fabrication constraints and analyze how they affect the device functionality. Specifically, we optimize for effective potentials which enable three devices: a device that generates a flat-top polariton distribution, a metalens that focuses a propagating polariton, and a nonlinearly activated isolator for a propagating polariton.\\

We now provide details on the iterative workflow used to optimize polariton devices. We first introduce the physical model, after which we outline the particulars of the forward simulation. Subsequently, we provide an overview of the adjoint method, topology optimization, and the tools we employed. Finally, we summarize the entire inverse design pipeline used in this work. The subsequent section showcases examples of designed structures.\\

We strive to optimize the effective potential landscape experienced by polariton condensates towards useful polaritonic devices. We model the dynamics of the polariton macroscopic wavefunction $\psi(x, y, t)$ with the open-dissipative Gross Pitaevskii (GPE) model \cite{kavokin2017microcavities}

\begin{align}\label{eqn:GPE_rot_frame}
    i \hbar \partial_t \psi =& \left[ - \frac{\hbar^2}{2m} \nabla^2 + V(x, y) + U |\psi|^2 - i \kappa \right] \psi \\ \nonumber
    &+ i P(x, y)\, ,
\end{align}

\noindent where the dependence of $\psi=\psi(x,y,t)$ is implicit. The first part of the GPE is a Schrödinger equation with $\hbar$ as Planck's constant, $m$ the polariton mass, and $V(x, y)$ the potential the exciton-polariton experiences. $U$ is the nonlinearity of the exciton-polariton condensate. Since the exciton-polariton condensate is typically not a closed system, a pump $P(x, y)$, and a linear decay rate $\kappa$ model the addition and loss of exciton-polaritons inside the system. Note that equation \ref{eqn:GPE_rot_frame} has been transformed to the rotating frame of the pump to avoid numerical instabilities due to high frequency oscillatory terms (see Supporting Information).
For the optimization, $\psi$ also functionally depends on $V(x, y)$, which is always implicitly assumed.\\

Topology optimization requires two simulations of our system. A forward simulation simulates the physical system, and a backward simulation calculates the gradients. The forward simulation evaluates the figure of merit, while the backward simulation calculates the gradients of the figure of merit with respect to the free parameters.\\
We then simulate the time evolution of any given polariton condensate inside our simulation domain.
As the simulation domain is finite, we have to choose our boundary conditions to avoid scattering at the edges, which would lead to nonsensical designs.
To avoid this issue, we implement perfectly matched layers (PML) for nonlinear Schrödinger equations \cite{antoine2020perfectly}.\\
To ensure the differentiability of our code, the entire simulation is done using the Google JAX framework \cite{jax2018github}, a software package that can automatically differentiate native Python and Numpy code. The ordinary differential solver is provided by diffrax \cite{kidger2021on}, a library of tools that can be used for automatic differentiation.\\

As we are interested in free-form optimization of our entire design space, every pixel of the design region is an optimization parameter. This results in tens of thousands of optimization parameters, making global optimization impossible. When looking at optimization problems of this size, gradient-based methods are typically used. While gradient-based optimization cannot guarantee a global minimum, it is very efficient for finding local optima inside huge parameter spaces. This is because the gradients are typically obtained using adjoint sensitivity analysis \cite{CAO2002171}. The main advantage of adjoint sensitivity analysis is that the cost of calculating the gradients is independent of the number of input parameters.\\ 
The topology optimization problem for our polariton condensate can be formulated as
\begin{align}
    \min_{V} \mathcal{L}(\psi(V), V)& \,\,\,\,\,\,\mathcal{L}: \mathbb{C}\times\mathcal{D} \rightarrow \mathbb{R}\\
    \text{s. t.~}\, i \hbar \partial_t \psi =& - \frac{\hbar^2}{2m} \nabla^2 \psi + V(x, y) \psi + U |\psi|^2 \psi\\ \nonumber
    &- i \kappa \psi + i P(x, y)\, . 
\end{align}
We want to minimize the figure of merit $\mathcal{L}$ by finding the optimal effective potential $V(x, y)$ inside the design region $\mathcal{D}$ under the constraint that the GPE holds. Given a discretized form of our macroscopic wavefunction and potential, we assume every pixel inside the design region is a free parameter for our optimization.\\
During each optimization step, the GPE is solved until the final timestep $T>0$ or until a steady state is reached, depending on the problem at hand. The solution $\psi(x, y, T)$ is then used to evaluate $\mathcal{L}$. Using the adjoint sensitivity method, we calculate the gradients $\frac{\partial \mathcal{L}}{\partial V}$.
We then use this gradient information to find a local minimum of $\mathcal{L}$ by using gradient-based optimization algorithms, such as the method of moving asymptotes \cite{svanberg1987method} and L-BFGS \cite{avriel2020nonlinear}\\
All two-dimensional potentials are simulated using an NVIDIA A100 Tensor-Core GPU. The one-dimensional simulations are done using an Intel(R) Core(TM) i7-10700T CPU.
In addition to that, we use a resolution of $\SI{40}{px}$ per $\mu m$, resulting in a parameter spaces of up to a few hundred thousand free parameters.
In total, a full optimization for our examples takes a few hours. More details can be found in the supplementary material.
In all of our optimizations we start with an initial potential $V=0$. In our rotating frame, this corresponds to a pump that is exactly resonant with the polaritons. The final optimized potentials have regions that are locally not in resonance, which could lead to additional dynamics. For example, an optical bistability could occur in regions that are close to resonance, potentially resulting in dynamical instabilities \cite{Baas2004}. Nonetheless, these dynamics are not the focus of our work and not observed in our examples.
\\

We present three selected designs to highlight the versatility of topology optimization of polaritonic devices. We design a potential that leads to a flat-top distribution for the polariton, a metalens that focuses an incident polariton, and a nonlinearly activated isolator.\\
We use a system of units where $\hbar=\SI{0.6582}{ps\cdot meV}$ and the electron mass is $m_e = \SI{6585}{meV\cdot ps^2\cdot \mu m^{-2}}$. The polariton mass is set to be $m_p = 10^{-4}\cdot m_e = \SI{0.6585}{meV\cdot ps^2\cdot \mu m^{-2}}$. Since most fabrication methods cannot fabricate arbitrarily large structures, we restrict the effective potential to a maximum difference of $\Delta V = \SI{50}{meV}$. Unless stated otherwise, we assume a polariton-polariton nonlinearity of $U = \SI{5}{neV\cdot \mu m^2}$, which corresponds to typical values observed in organic microcavities \cite{Keeling2020Boseeinstein}. We choose the linear loss $\kappa$ according to the problem discussed.

\begin{figure*}
    \centering
    \includegraphics[width=0.8\textwidth]{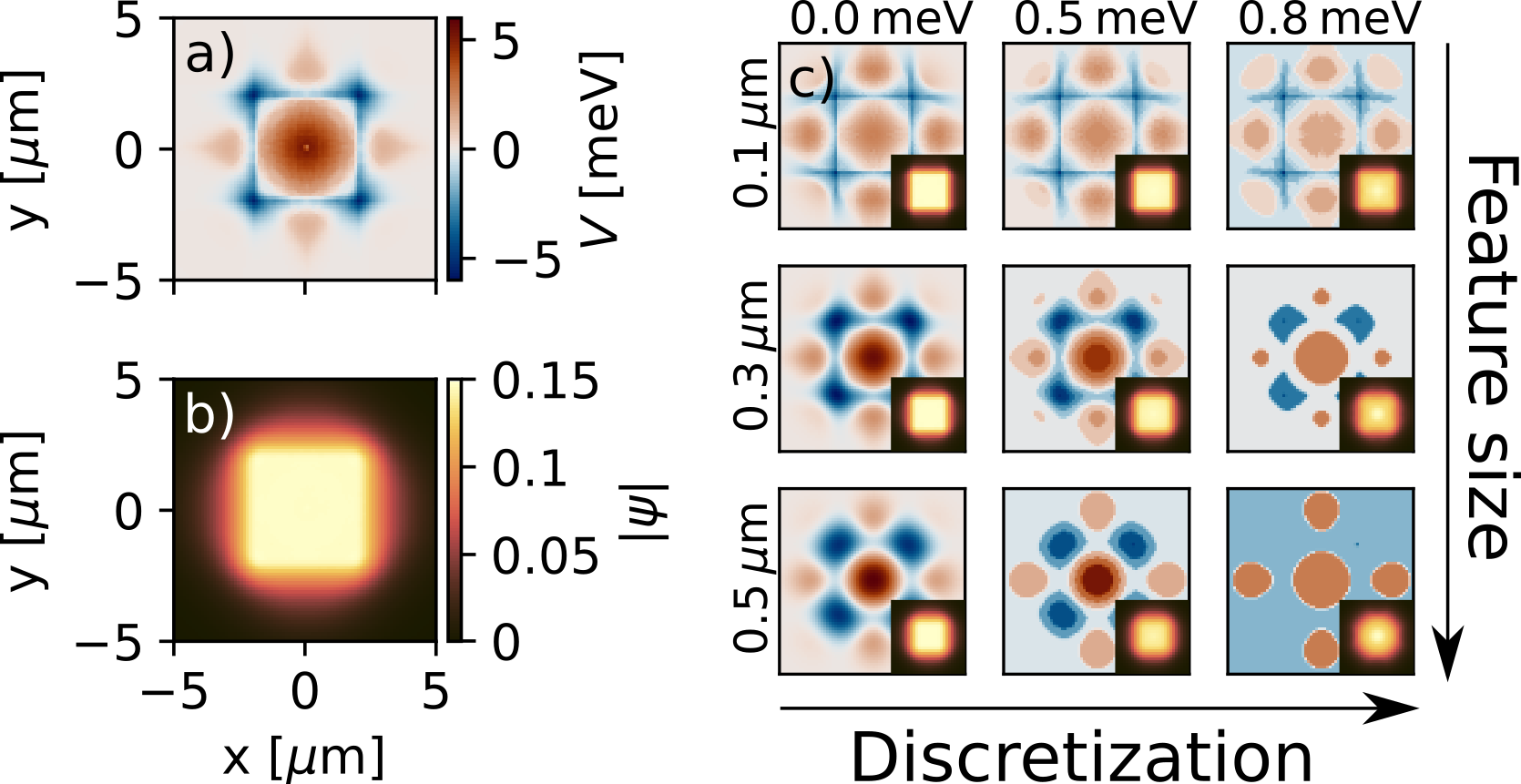}
    \caption{Optimization results for the polaritonic flat top potential. a) The effective potential the exciton-polariton condensate experiences in the rotating frame. b) The distribution of the exciton-polariton condensate in the steady state. c) The effect of imposing a minimum feature size on the potential (row) and the effect of discretization of the potential (column). The respective insets show the associated exciton-polariton condensate distribution of the potentials.}
    \label{fig:flat_top}
\end{figure*}
At first, we present the optimized potential, which gives rise to a flat-top distribution.
Having a square flat top is of big interest in optics. A constant irradiation profile is a desired feature when it comes to semiconductor fabrication, material heating, and meteorology. This idea can be extended to exciton-polariton condensates. Upon illumination by a typical laser, the exciton-polariton condensate is spatially distributed according to a Gaussian distribution if no further measures are implemented. A pump
\begin{equation}
    P(x, y) = e^{-\frac{x^2 + y^2}{\SI{2}{\micro \meter}^2}}
\end{equation}
is used to simulate a constant, resonant illumination at normal incidence.
To model the combined radiative and non-radiative losses in the condensate, a decay rate of $\kappa=\SI{4}{meV}$ is used. This allows the exciton-polariton condensate to reach a steady state after around $\SI{3}{ps}$. \\
To achieve a flat top, we choose a figure of merit that ensures vanishing spatial gradients of the condensate wavefunction within the design region. Specifically, we minimize the figure of merit
\begin{equation}
    \mathcal{L}(\psi(V), V) = \sum\limits_{(x, y)\in \mathcal{D}} |\nabla \psi(x, y)|\, .
\end{equation}
Here, $\mathcal{D}\subseteq \mathbb{R}$ represents the flat top region.
As the gradients outside $\mathcal{D}$ are mostly zero, the design region naturally confines itself to $\mathcal{D}$.
The entire simulation domain has a size of $\SI{20}{\mu m}\times \SI{20}{\mu m}$.
The resulting potential and the associated structure can be seen in \autoref{fig:flat_top} a) and b).
Note that we do not show the entire simulation domain, as wavefunction and potential are zero everywhere else. These potentials have been optimized for a flat top of size $\SI{2}{\mu m} \times \SI{2}{\mu m}$.

Different minimum feature sizes are imposed during the optimization to accommodate different experimental conditions \cite{Schneider2017Exciton}, e.g. the resolution of fabrication techniques such as focused ion beam milling \cite{Urbonas2016}. We note that optically created potentials \cite{Askitopoulos2013} often have an even larger minimum feature size than considered here (a few micrometers), which is related to the healing length of the polariton fluid.
We consider two restrictions: The minimum feature size of the spatial features of the potential $\Delta x$ and a minimal step size of the values of the potential itself $\Delta V(x, y)$.
A Gaussian blur is applied to the potential at every optimization step for the minimum feature size. This forces the optimization algorithm to find solutions above a specific feature size. This feature size corresponds roughly to $\sqrt{3}\cdot \sigma_G$, where $\sigma_G$ is the standard deviation of the Gaussian blur.
We do not enforce a minimal step size of the potential during the optimization loop. This is because it is difficult to implement the discretization in a continuously differentiable manner, which is necessary for calculating the gradients. Instead, we optimize the potential without any constraints (starting from $V=0$) and then apply discretization to the final optimized potential for various different feature sizes. This is done by taking multiples of $\Delta V$ and assigning the values of the potential to the closest discretized level. This results in a discretized potential, which we then use to solve the GPE in a single forward solve. The different potentials are shown in \autoref{fig:flat_top} c) with their respective exciton-polariton distribution in the inset.\\
The main standout features for small discretizations are the circular center of positive potential and the surrounding square made out of negative potential. The thickness of the square mentioned above determines the sharpness of the edges of the exciton-polariton flat top. Increasing the minimum feature size results in less sharp edges of the flat top. Still, the potential can produce a flat top distribution even at a minimum feature size of around $\SI{0.5}{\mu m}$.
Increasing the discretization has a different effect. The square in the potential becomes increasingly disconnected until only the most prominent features remain. Increasing the discretization level leads to a loss in the flatness of the flat top distribution, and the exciton-polariton condensate becomes more Gaussian-shaped.\\

\begin{figure}
    \centering
    \includegraphics[width=0.4\textwidth]{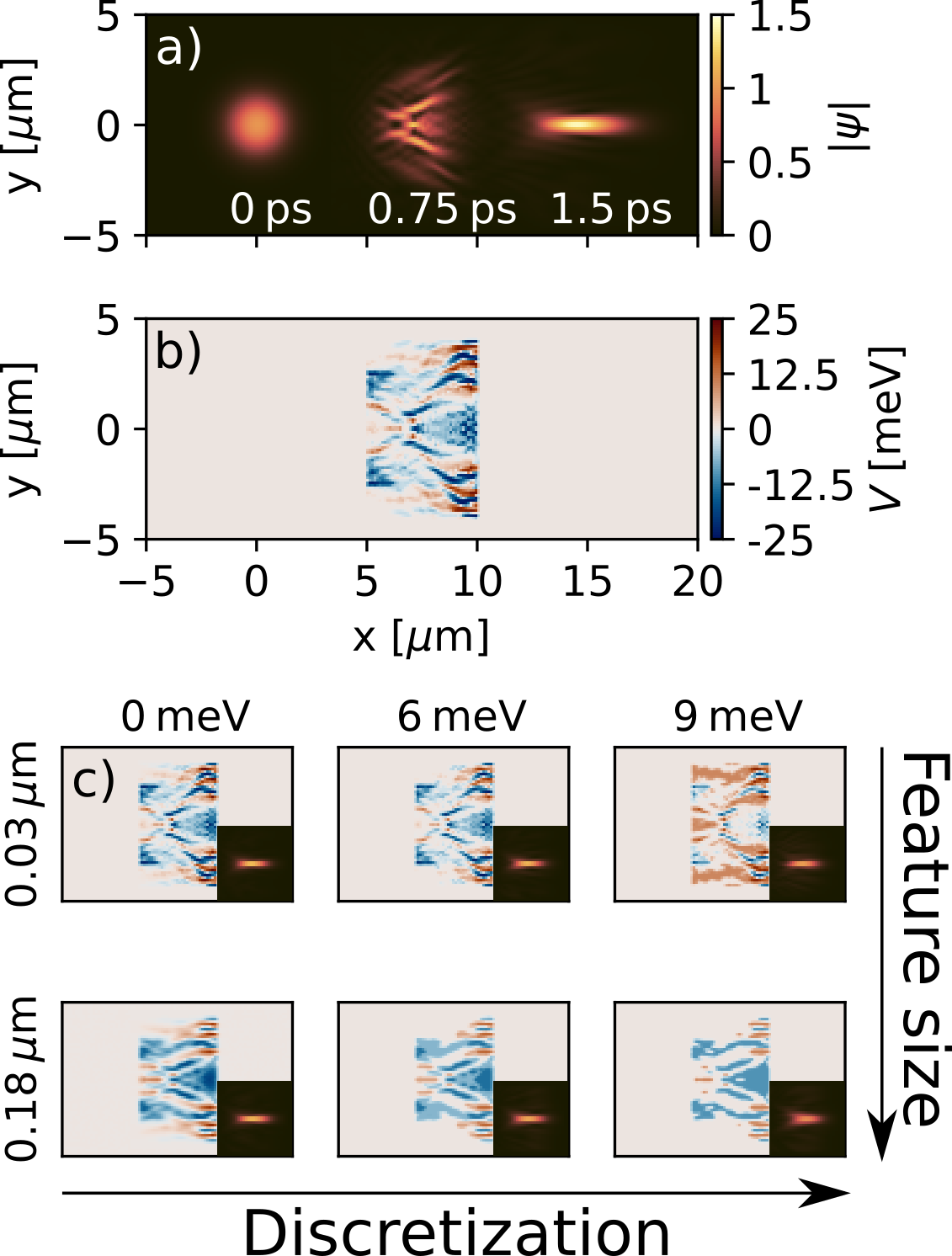}
    \caption{Optimization results for the polaritonic lens. a) The exciton-polariton condensate distribution for three different timesteps. b) The effective potential of the exciton-polariton experiences in the rotating frame. c) The effects of imposing a minimum feature size on the potential (row) and the effect of discretization of the potential (column). The respective insets show the associated exciton-polariton condensate distribution.}
    \label{fig:lens}
\end{figure}

Lenses are some of the most fundamental and useful optical devices. Metalenses, in particular, find applications in optoelectronics and 3D imaging. For exciton-polariton condensates, lenses could be useful components in optoelectronic circuits.\\
Instead of considering a steady state solution like the flat top, we optimize a metalens for a propagating polariton packet. No pump is used, and the system is simplified with $\kappa=0$. The initial wave packet is initialized as
\begin{equation}
    \psi_0(x, y) = e^{-\frac{x^2 + y^2}{\SI{2}{\mu m}^2}} e^{i\SI{10}{\mu m^{-1}} x}\, .
\end{equation}
The wave packet is propagated for $\SI{1.5}{ps}$ according to the GPE. The figure of merit
\begin{equation}
    \mathcal{L}(\psi) = |\psi(x_0, y_0)|\,\,\, (x_0, y_0) \in \mathbb{R}
\end{equation}
tries to maximize the value at the focal point $(x_0, y_0)$ at $T=\SI{1.5}{ps}$. An explicit design region of size $\SI{5}{\mu m} \times \SI{8}{\mu m}$ is specified. The wave packet propagates for $\SI{5}{\mu m}$ until it enters the design region. The focal point is at $(x_0, y_0) = (\SI{15}{\mu m}, \SI{0}{\mu m}$), $\SI{5}{\mu m}$ behind the design region. 
The evolution of the wave packet can be seen in \autoref{fig:lens} a). The initial wave packet is Gaussian-shaped and evolved according to the GPE. As the wave packet propagates in positive $x$-direction, it passes the design region, where it experiences the effective potential seen in \autoref{fig:lens} b). The effective potential acts as a metalens, with its focal point behind the effective potential. 
The lens can focus the exciton-polariton condensate at a tight spot and reach an enhancement of up to $1.5$ compared to a condensate without potential.\\
Again, a minimum feature size is imposed during optimization to accommodate realistic fabrication restrictions, and the optimized potential is discretized. The resulting structures for different minimum feature sizes and discretizations can be seen in \autoref{fig:lens} c).
The metalens uses channels of negative potential to guide the polariton-condensate through the design region. The enhancement is achieved by having the entire wave packet redirected towards the focal point, similar to an optical metalens. Imposing fabrication restrictions has little effect on the functionality of the polaritonic metalens, at least for realistic fabrication restrictions.\\

\begin{figure}
    \centering
    \includegraphics[width=0.4\textwidth]{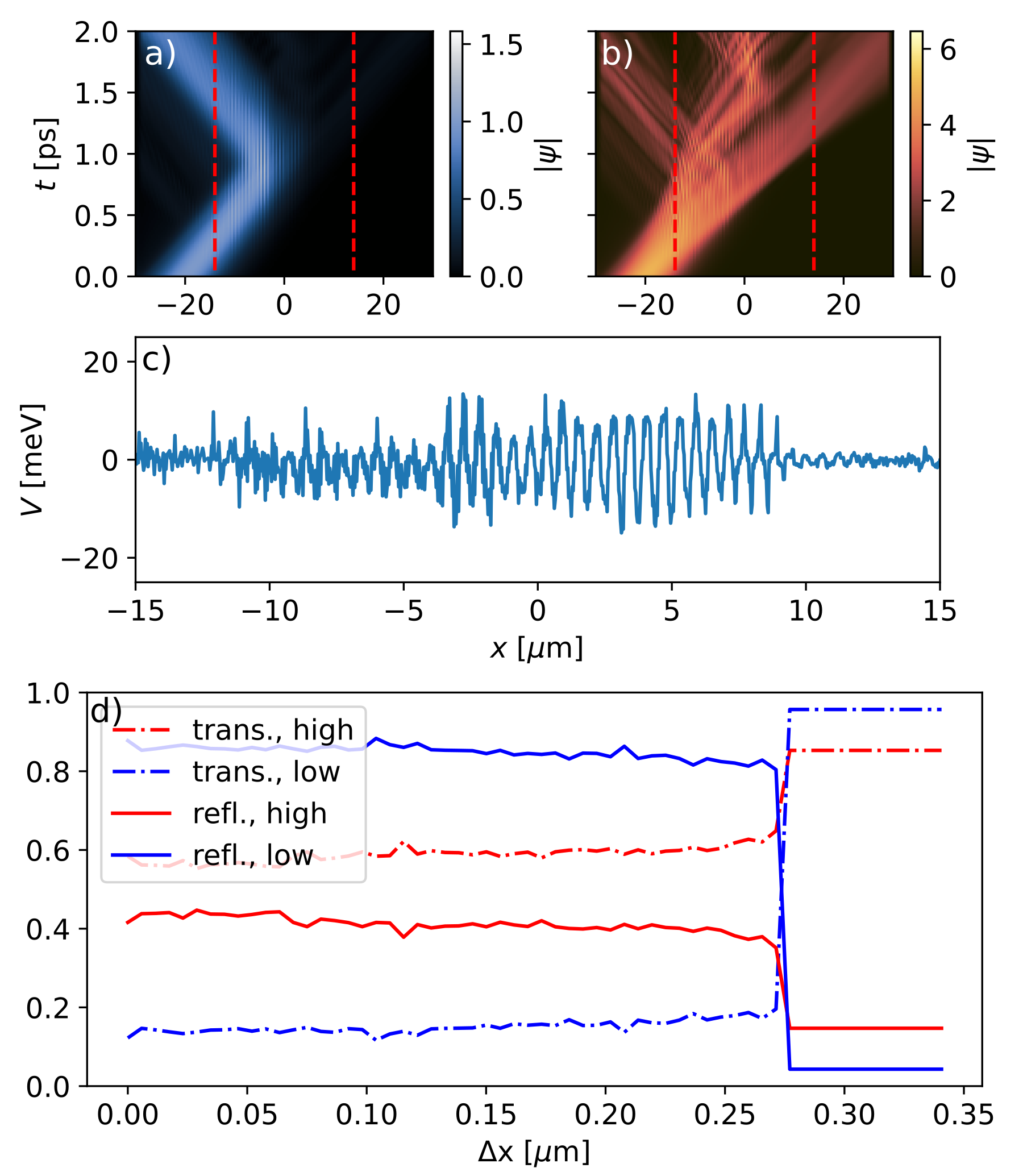}
    \caption{Optimization results for the polaritonic mirror. a) The evolution of the low amplitude exciton-polariton condensate in time. The red dashed lines indicate the boundaries of the potential. b) The evolution of the high amplitude exciton-polariton condensate in time. The red dashed lines indicate the boundaries of the potential. c) The effective potential the exciton-polariton condensate experiences in the rotating frame. d) Transmissivity of the high amplitude exciton-polariton condensate (red, dash-dotted), transmissivity of the low amplitude exciton-polariton condensate (blue, dash-dotted), reflectivity of the high amplitude exciton-polariton condensate (red, solid), and reflectivity of the low amplitude exciton-polariton condensate (blue, solid) depending on the minimum feature size imposed.}
    \label{fig:mirror}
\end{figure}

So far, the nonlinearity of the GPE has been neglected as a design parameter. Especially in the context of nanophotonics, nonlinear optics shows many promising results. On the other hand, nonlinear exciton-polariton condensates are being studied for their possible applications in optoelectronics and optical and quantum computing.
A nonlinearly activated isolator is designed to exploit the nonlinearity of the polariton condensate. A one-dimensional wave packet is propagated for $\SI{4}{ps}$ in positive $x$-direction. The initial wave packet is set to
\begin{equation}
    \psi_0^\mathrm{h/l}(x, y) = A^\mathrm{h/l} e^{-\frac{(x-\SI{20}{\mu m})^2}{2\cdot\SI{5}{\mu m}^2}} e^{i \SI{10}{\mu m}^{-1}\cdot x}\, ,
\end{equation}
where $\mathrm{h}$ (resp. $\mathrm{l}$) denotes a "high" (resp. "low'') amplitude wave packet. The nonlinearity is set to $U=\SI{0.5}{meV \mu m^2}$. Such high nonlinearities can be achieved with Perovskite materials \cite{Fieramosca2019}. The nonlinear mirror is optimized to be reflective for the low amplitude exciton-polariton condensate and transmissive for the high amplitude exciton-polariton condensate. The figure of merit is
\begin{equation}
    \mathcal{L}(\psi) = \frac{\sum\limits_{x<\mathcal{D}} |\psi^\mathrm{l}(x)|^2}{\sum\limits_x |\psi^\mathrm{l}(x)|^2} + \frac{\sum\limits_{x\ge\mathcal{D}} |\psi^\mathrm{h}(x)|^2}{\sum\limits_x |\psi^\mathrm{h}(x)|^2}\, .
\end{equation}
Here, $\mathcal{D}$ represents the left border of the design region. The one-dimensional structure is optimized for the two amplitudes $A^\mathrm{l}=0.5$ and $A^\mathrm{h}=5$ and restricted to a design region of size $\SI{30}{\mu m}$. 
The evolution of the exciton-polariton condensate distribution and the respective nonlinear mirror can be seen in \autoref{fig:mirror} a)-c). The potential itself varies strongly on a short-length scale. The potential acts like a Bragg mirror for the low amplitude exciton-polariton condensate. For the high amplitude exciton-polariton condensate, the nonlinearity, and by extension, the interaction of the exciton-polaritons, causes the exciton-polariton condensate to scatter in a way that allows for the exciton-polariton condensate to be transmitted through the nonlinear mirror partially. The nonlinear mirror can achieve a reflectivity of around $0.9$ for the low amplitude exciton-polariton condensate and a transmission of around $0.4$ for the high amplitude exciton-polariton condensate.

As with the previous two designs, a minimum feature size $\Delta x$ is implemented during optimization. The resulting reflectivity and transmissivity for the high and low amplitude exciton-polariton condensate can be seen in \autoref{fig:mirror} d). The performance of the nonlinear mirror stays practically the same until around $\Delta x \approx \SI{0.28}{\mu m}$. After that, a rapid drop in functionality occurs, as the nonlinear mirror does not function at all after this point. This can be attributed to the exciton-polariton condensate having an inherited wavelength, which the nonlinear mirror is optimized for. Once the spacing between the peaks of the potential becomes too large, the functionality breaks down as the exciton-polariton condensate cannot be adequately scattered anymore.\\

In conclusion, we described an inverse design approach to optimize the potential that governs the propagation characteristics of a polariton condensate to implement a set of functional devices with increasing complexity. Our approach is particularly appealing, as the optimized potential can frequently be explicitly controlled in an experiment. The topology optimization allows us to accommodate experimental constraints such as a minimal feature size or a discretization of the values it can attain. Of course, being more restrictive causes a degradation of the objection function. Still, ultimately, it is an engineering question of how much effort can be spent to fabricate a given device to keep the possible restriction in the fabrication to a minimum.

We demonstrate the optimization pipelines on three devices with increasing complexity. First, we consider a steady-state situation. We designed a polariton condensate with a flat-top distribution for a given Gaussian pump. Second, we consider a propagating polariton condensate that we localize at a predefined spatial and temporal location. Third, we consider an explicitly nonlinear device that reflects the polariton at a low amplitude but transmits it at a high amplitude. We consistently elaborated on the impact of the minimal feature size and the discretization of the potential on the achievable functionality. While generally a degradation is encountered, the designed devices are robust against such limitations and uphold the desired functionality. A sudden functionality degradation was observed only for the last device when the typical feature size reached a threshold. The finding was explained by observing that the necessary Bragg-like feature could no longer be provided above the observed threshold.

To build on our findings, future developments in the inverse design of polariton condensates could explore the integration of machine learning algorithms to enhance the optimization process. Leveraging advanced computational techniques makes it conceivable to predict and design potentials with unprecedented precision. The potential applications of this inverse design approach are vast and promising. The ability to control polariton condensates could significantly impact the development of various applications, from highly efficient, tunable photonic devices and quantum information processing components to innovative approaches in sensing and imaging technologies, significantly impacting fields such as telecommunications and medical diagnostics.

\begin{acknowledgments}
O.K., Y.A., and C.R. acknowledge support from the German Research Foundation within the Excellence Cluster 3D Matter Made to Order (EXC 2082/1 under project number 390761711) and by the Carl Zeiss Foundation.
T.J.S. acknowledges funding from the Alexander von Humboldt Foundation. 
\end{acknowledgments}

\section*{Data Availability Statement}
The data that support the findings of this study are available from the corresponding author upon reasonable request.
However, the code is open source and available under \url{https://gitlab.kit.edu/oliver.kuster/inverse_design_of_polaritonic_devices}. The code can be used to reproduce the data and plots presented in this work.

\nocite{*}
\bibliography{Main_ArticleAPL}

\section*{Supplementary material}

\subsection{Rotating frame}

In the main text we use the GPE in a rotating frame. The rotating frame is obtained from the initial GPE

\begin{equation}\label{eqn:GPE_not_rot}
    i \hbar \partial_t \phi = \hat{H}_\text{L}\phi + U |\phi|^2 \phi + i F(x,y,t),
\end{equation}

where $\phi \equiv \phi(x, y, t)$ is the condensate wavefunction and $U$ is the strength of the nonlinearity. The terms linear in $\phi$ are

\begin{equation}
    \hat{H}_L = -\frac{\hbar^2}{2m} \nabla^2 + \mathcal{V}(x, y) - i \kappa,
\end{equation}

where $\kappa$ is the polariton loss rate, $m$ is the effective mass, and $\mathcal{V}$ is the effective potential. In the main text we assume that the pump is a harmonic function of the form $F(x,y,t)=P(x,y)\exp(-i\omega_P t)$. Let us define the condensate wavefunction in a rotating frame as $\psi(x,y,t)=\phi(x,y,t) \exp(i \omega_P t)$. Upon multiplication by $\exp(i \omega_P t)$, the following transformation occur: $\hat{H}_L \phi \mapsto \hat{H}_L \psi$, $|\phi|^2 \phi \mapsto |\psi|^2 \psi$, $F(x,y,t) \mapsto P(x, y)$, and $i \partial_t \phi \mapsto \omega_P \psi + i \partial_t \psi$. Therefore, in the rotating frame, equation \ref{eqn:GPE_not_rot} becomes

\begin{equation}
    i \partial_t \psi = \left[ -\frac{\hbar}{2m} \nabla^2 + V(x, y) - i \kappa \right] \psi + U |\psi|^2 \psi + i P(x,y),
\end{equation}

where we have introduced the renormalised effective potential $V(x, y) = \mathcal{V}(x, y) - \hbar \omega_P$.

\subsection{Numerical details and benchmarks}
As mentioned, the two-dimensional potentials are simulated using an NVIDIA A100 Tensor-Core GPU and the one-dimensional simulations are done using an Intel(R) Core(TM) i7-10700T CPU. The required optimization steps and optimization time vary depending on the complexity of the structure.
We use a spatial resolution of $\SI{40}{px}$ per micrometer, while the timestepping is done by using a PID controller \cite{9780134685717}.
Using this setup the simulations tend to reach convergence around 50-150 optimization steps. 
We will not go into more detail for the one dimensional case, as the simulation itself is fairly fast and converges in less than an hour on almost any hardware.\\
\begin{figure}
    \centering
    \includegraphics[width=\linewidth]{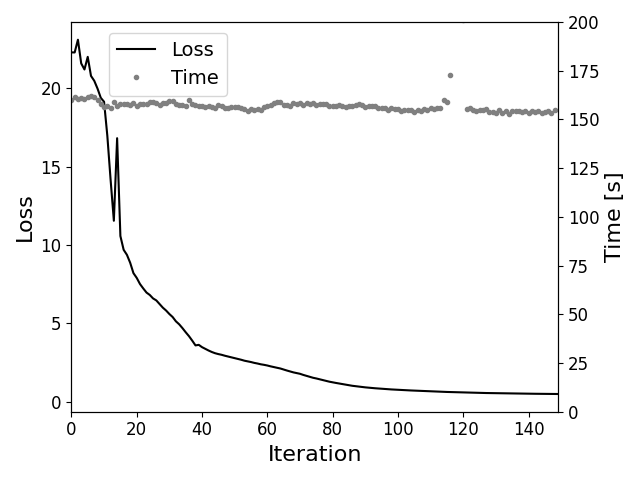}
    \caption{Loss graph for the optimization of a flat top. The dots show the time required at each optimization step, resulting in a total optimization time of roughly $\SI{6.5}{h}$ and an average time of $\SI{159}{s}$ per iteration.}
    \label{fig:flat_loss}
\end{figure}
\begin{figure}
    \centering
    \includegraphics[width=\linewidth]{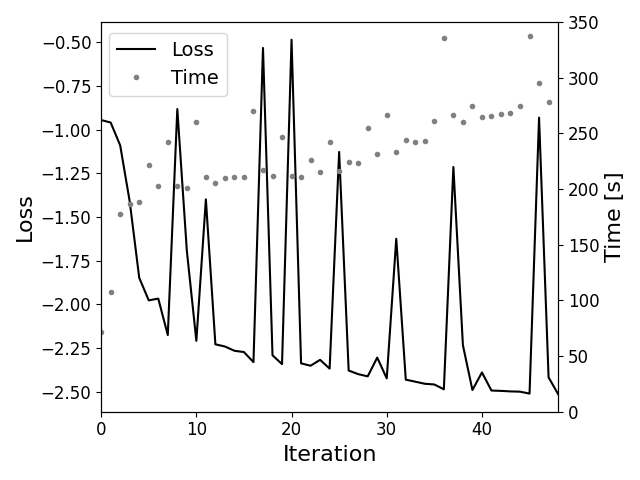}
    \caption{Loss graph for the optimization of a lens. The dots show the time required at each optimization step, resulting in a total optimization time of roughly $\SI{3}{h}$ and an average time of $\SI{233}{s}$ per iteration.}
    \label{fig:lens_loss}
\end{figure}

Both flat top and lens use Runge Kutta methods, specifically Tsitouras' 5/4 method \cite{tsitouras2011runge}, for the evolution of the GPE in time. The main difference between those two simulations is that the flat top is simulated until a steady state is reached, while the lens is simulated until $t=\SI{1.5}{ps}$ is reached.
The specific parameters which define step size and steady state criteria can be set manually in the simulation and are provided by the package which implements the solver.
We also decrease the number of max steps for the lens, as the system is bigger and would otherwise run out of memory. Alternatively, checkpointing in the adjoint simulation can be used to reduce the memory demand.\\
Both systems are optimized until either $150$ optimization steps or convergence is reached. The loss and time required for every epoch can be seen in \autoref{fig:flat_loss} for the flat top and \autoref{fig:lens_loss} for the lens.
The optimization of the flat top runs for $150$ iterations and takes approximately $\SI{6.5}{h}$. Each iteration for the optimization takes roughly $\SI{159}{s}$ and the optimization is best done with L-BFGS.
The optimization of the lens reaches convergence after approximately $\SI{3}{h}$ and $48$ iterations. Per iteration, one optimization step took roughly $\SI{233}{s}$. For the most part L-BFGS and MMA tend to perform similarly for the lens.

\end{document}